\documentclass[onecolumn,superscriptaddress,floats,floatfix,aps,pra]{revtex4-2}

\usepackage{bm} 
\usepackage{latexsym}
\usepackage{xcolor}
\usepackage{graphicx}
\usepackage{hyperref}

\usepackage{url}
\usepackage{enumerate}
\usepackage{MnSymbol}
\usepackage{bbding}
\usepackage{float}
\usepackage{soul}
\usepackage{tikz}
\usetikzlibrary{shapes}
\usepackage{wrapfig}
\usepackage{ulem}

\definecolor{burgundy}{rgb}{0.5, 0.0, 0.13}

\newcommand{\w}{\omega}
\newcommand{\tw}{\text{w}}
\newcommand{\eps}{\epsilon}
\newcommand{\g}{\gamma}

\renewcommand{\a}{\hat{a}}
\newcommand{\ac}{\hat{a}^\dagger}
\newcommand{\op}{\hat{o}}
\newcommand{\opc}{\hat{o}^\dagger}
\renewcommand{\b}{\hat{b}}
\newcommand{\bc}{\hat{b}^\dagger}
\renewcommand{\c}{\hat{c}}
\newcommand{\cc}{\hat{c}^\dagger}
\newcommand{\rh}{\hat{\rho}}

\newcommand{\intw}{\int^{\infty}_{-\infty}d\w\;}

\newcommand{\av}[1]{\langle{#1}\rangle}

\newcommand{\Pearson}[1]{{\cal C}_{#1}}

\newcommand{\figref}[1]{fig.(\ref{#1})}

\begin{document}
	\bibliographystyle{unsrt}
	
	\title{Quantum  synchronisation and clustering in chiral networks}
	
	\author{Salvatore Lorenzo}
	\affiliation{Universit{\`a}  degli Studi di Palermo, Dipartimento di Fisica e Chimica -- Emilio Segr{\`e}, via Archirafi 36, I-90123 Palermo, Italy}
	
	\author{Benedetto Militello} 
	\author{Anna Napoli}
	\affiliation{Universit{\`a}  degli Studi di Palermo, Dipartimento di Fisica e Chimica -- Emilio Segr{\`e}, via Archirafi 36, I-90123 Palermo, Italy}
	\affiliation{INFN Sezione di Catania, via Santa Sofia 64, I-95123 Catania, Italy}
	
	\author{ Roberta Zambrini} 
	\affiliation{Instituto de F\'isica Interdisciplinar y Sistemas Complejos (IFISC, UIB-CSIC) Campus Universitat de les Illes Balears, E-07122 Palma de Mallorca, Spain}
	
	\author{ G.Massimo Palma}
	\affiliation{Universit{\`a}  degli Studi di Palermo, Dipartimento di Fisica e Chimica -- Emilio Segr{\`e}, via Archirafi 36, I-90123 Palermo, Italy}
    \affiliation{NEST, Istituto Nanoscienze-CNR, Piazza S. Silvestro 12, 56127 Pisa, Italy}

\begin{abstract}
We study the emergence of synchronisation in a chiral network of harmonic oscillators. The network consists of a set of locally incoherently pumped harmonic oscillators coupled pairwise in cascade with travelling field modes. 
Such cascaded coupling leads to feedback-less dissipative interaction between the harmonic oscillators of the pair which can be described in terms of an effective pairwise hamiltonian a collective pairwise decay. 
The network is described mathematically in terms of a directed graph. By analysing geometries of increasing complexity we show how the onset of synchronisation depends strongly on the network topology, with the emergence of synchronised communities in the case of complex networks. 
The quantum nature of the non local correlation between network nodes is assessed.
\end{abstract}

\maketitle

\section{Introduction}
Spontanoeus synchronisation can be described as the emergence of a collective dynamics in a set of interacting subsystems such that all the individual parts of the whole system evolve in the same way in spite of their differences~\cite{pikovsky2001}.
This class of phenomena is relevant in several contexts: from bridge engineering~\cite{refEckhardt} to neurosciences~\cite{refStrogatz,refAngelini}, from neural networks~\cite{refLodi} to synchronized motion of butterflies and fireflies~\cite{refPardikes} and also in social~\cite{neda2000} and chemical~\cite{Taylor2009} systems.
If an external forcing is applied, the ability of the system to follow the driver dynamics defines instead the emergence of driven synchronization~\cite{pikovsky2001}.
In the last decades, attention to synchronisation phenomena has moved from the realm of classical mechanics~\cite{refAcebron,refMaianti,refArenas,refPantaleone} to the quantum physics one~\cite{Hanggi2006,Zhirov2008,Galve2012,Lee2013,refMendoza,galve2017}, with particular interest for the behaviour of networks of interacting quantum subsystems~\cite{manzano2013,cabot2018}.

Furthermore the connection between synchronisation and other collective phenomena such as superradiance and subradiance~\cite{refBellomo} as well as the synchronisation of hybrid quantum systems consisting of coupled quantum oscillators and few-state systems has  been analysed~\cite{refHush,refMilitello}.
While the emergence of synchronisation is generally linked to the interplay between dissipative dynamics, nonlinear couplings and driving, in the case of quantum systems an important role is played also by quantum noise and quantum correlations~\cite{galve2017}.

In this work we will analyse the emergence of synchronisation patterns in networks of coupled quantum harmonic oscillators~\cite{manzano2013,cabot2018}
in which the effective interaction between subsystems is due to a chiral coupling to travelling modes.
The very rapid progress in the new field of chiral quantum optics~\cite{lodahl2017}, paves the way  for new ways to manipulate and control light-matter interaction. For instance the strong light confinement in nanophotonic structures~\cite{refPetersen,refMitsch}
can lead to propagation-direction-dependent emission, scattering and absorption of photons by quantum emitters inducing a propagation-direction-dependent light-matter interaction~\cite{refKornovan,refColes}. Within the framework of chiral quantum optics, it is also possible to realize cascaded quantum systems~\cite{lodahl2017,gardiner1985,gardiner1993,carmichael1993a} that could be exploited, for example, to drive synchronisation at demand.

The way in which the asymmetric role of the network nodes in the presence of a chiral coupling affects synchronisation is so-far unexplored.  
This has led us to analyse the emergence of synchronisation in a network of harmonic oscillators (HOs henceforth) connected via a one-directional coupling to travelling modes. Furthermore incoherent coupling is also considered allowing for a sustained dynamics, going beyond transient synchronization~\cite{transientbookchapter}.

We have found a strong dependence of the synchronisation patterns on the network size and topology. In particular, for networks involving a high number of nodes and complex topology we predict the appearance of clusters of synchronised nodes with different cluster frequencies. In the next section we will introduce our model and its Hamiltonian and we obtain the equation of motion of the local oscillators when coupled to one-directional travelling modes via a master equation for the cascaded quantum systems. In sec.~\ref{sec:synchronisation} we introduce the quantitative witnesses we use to characterize the emergence of synchronisation. In sec.~\ref{sec:simplegraph} we report our main results, namely we analyse the emergence of different synchronisation patterns in networks of increasing complexity. In sec.~\ref{sec:conclusion} we sum-up of the previous sections and discuss future directions.


\section{Dynamics of chiral harmonic quantum networks}
\label{sec:system}
\begin{figure}[!ht]
	\centering
	\includegraphics[width=0.8\linewidth]{ 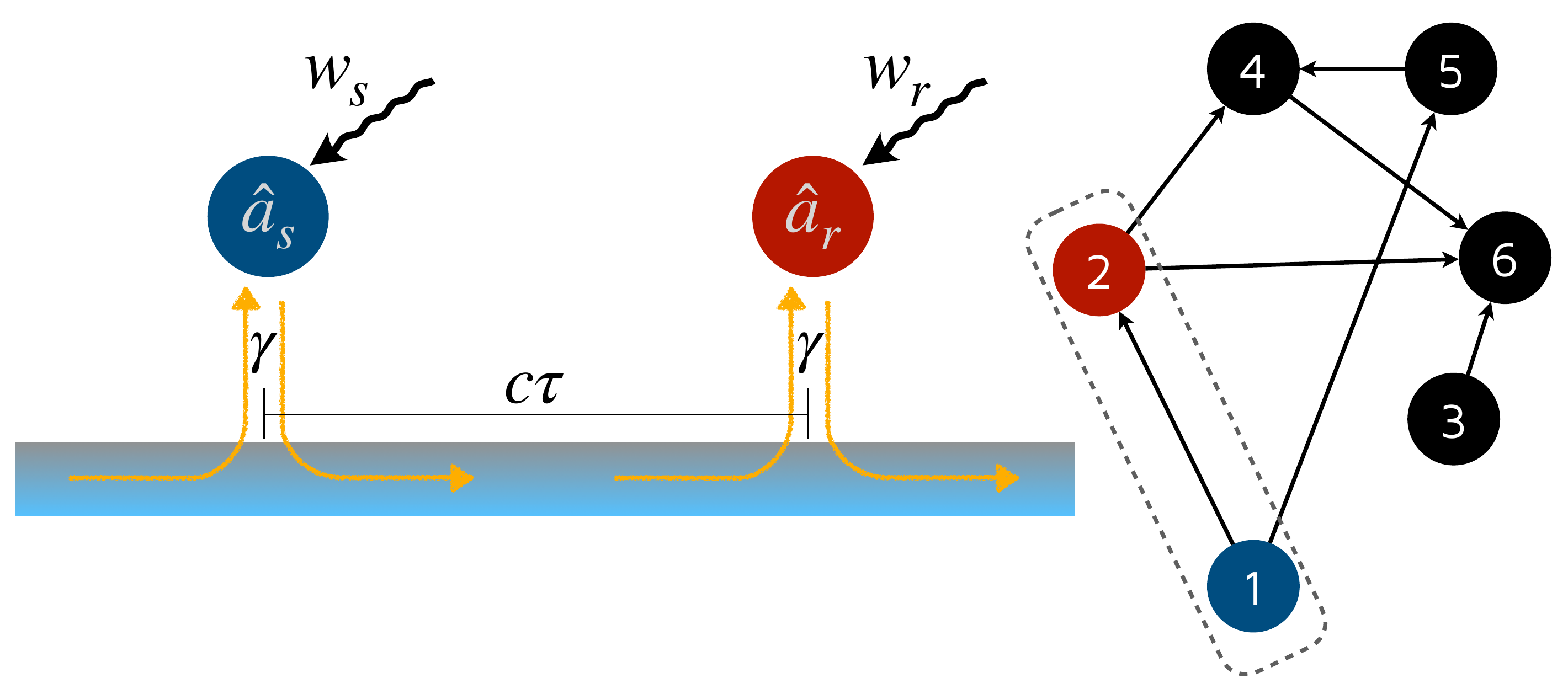}
	\caption{Sketch of pairwise coupling between nodes. A pair of HOs (blue and red, left) is coupled in cascade to travelling modes (orange lines). The cascaded coupling leads not only to dissipation but also to a chiral coupling in which the blue HO drives the red one but there is no feedback from the red to the blue one. Such coupling can described in terms of a pairwise effective interaction plus a pairwise collective decay, both of dissipative origin. In the right an example of direct graph in which each links represent exactly the pairwise coupling sketched in the left.}
	\label{fig:fig1}
\end{figure}

Our system consists of a chiral network of coupled harmonic oscillators which can be conveniently described in terms of a a directed graph, with adjacency matrix $\bf A$ (i.e. a set of nodes linked by directed edges), where each node corresponds to a single, incoherently pumped,  harmonic oscillator while each directed edge corresponds to a unidirectional cascaded coupling between the oscillators. As we will consider strictly cascaded systems~\cite{lodahl2017,gardiner1985,gardiner1993,carmichael1993a,giovannetti2012a} we  restrict our attention to oriented graphs, i.e. we assume that each pair of nodes is linked by a single oriented edge ($A_{ij}=1$ implies $A_{ji}=0$). The cascaded coupling between a pair of oscillators $s,r$  \figref{fig:fig1}  is due to travelling field modes propagating from node $s$ to node $r$ ~\cite{wang2005,koch2010,kamal2011,feng2011} such that when a photon is emitted by $s$ it can only propagate towards $r$, while the reverse is not possible~\cite{gardiner1993,stannigel2012}. 
The dynamics of the network is generated by the sum of the node hamiltonians $H_{N}^{k}$, which includes the local energies and the incoherent pump on the oscillator  $k$,  and of the edge hamiltonians $H_{E}^{sr}$, which describes the unidirectional coupling between pairs of oscillators $s,r$ due to travelling modes, and can be written in a compact form as
\begin{align}
	\hat{H}= \sum_{\text{nodes}} {\hat H}_{N}^k + \sum_{\text{edges}} {\hat H}_E^{sr}\,.
\end{align}
For a pair of nodes the two terms read
\begin{align}\label{H}
H_{N} \,= &\sum_{k={s,r}}\;\eps_k\;\ac_k\a_k -\int_0^{\infty}d\omega\;\w\; \cc_{k\w}\c_{k\w} + \sqrt{\frac{\tw_k}{2\pi}}\intw \left(\a_k\c_{k\w}+h.c.\right) \\ 
	H_{E}^{sr}=& \int_0^{\infty} d\omega \;\w\; \bc_\w\b_\w +\sqrt{\frac{\g}{2\pi}}\int_0^{\infty} d\omega\; \left((\a_s+\a_re^{-i\w \tau})\bc_\w+h.c.\right)\nonumber
\end{align}
where each oscillator, with energy $\eps_k$ ($\hbar =1$) and bosonic creation and annihilation operators 
$\ac_k$ and $\a_k$, is incoherently pumped at a rate $\tw_k$ by a local bath of inverted harmonic oscillators with operators $\cc_k$ and $\c_k$~\cite{Glauber_Amp}. In the same way, operators $\bc_\w$ and  $\b_\w$ represent the travelling modes responsible of the coupling between nodes $s$ and $r$, and $\tau$ is the time it takes the the field to propagate from $s$ to $r$ (note that for the travelling modes  $\omega$ plays also the role of a wavevector and  the correct propagation directionality is ensured)~\cite{gardiner1993,stannigel2012}. 
In writing~\eqref{H} we have made both the {\it rotating wave} and the {\it Markov approximation}. The first consists in neglecting the rapidly  oscillating counter rotating terms in the interaction hamiltonian~\cite{harocheExploring2006,breuerTheory2007} while the second  assumes the coupling amplitudes, $\sqrt{\g/2\pi}$ and $\sqrt{\tw_k/2\pi}$ in~\eqref{H}, to be constant across a broad band around the characteristic frequencies of the local oscillators~\cite{gardiner2004a,wiseman2009,jacobs2014}.  

From~\eqref{H} it is immediate to derive the Heisenberg equations for the field operators:
\begin{align}
	&\frac {d \b_\w(t)} {d t}=-i\w \b_\w(t)-i\sqrt{\frac{\g}{2\pi}}\; (\a_s(t)+\a_r(t) e^{-i\w \tau})\label{bcEQS}\nonumber\\
	&\frac {d \c^\dagger_{k\w}(t)}{dt}=-i\w\c^\dagger_{k\w}(t)+i\sqrt{\frac{\tw_k}{2\pi}}\; \a_k(t)
\end{align}
and for the local oscillators operators $\a_r$,$\a_s$
\begin{align}
		\frac {d \a_s(t)}{dt}=
		&-i\eps_s\a_s(t)-i\sqrt{\frac{\g}{2\pi}}\int^{\infty}_{-\infty}d\w \b_\w(t)-i\sqrt{\frac{\tw_r}{2\pi}}\int^{\infty}_{-\infty}d\w \c^\dagger_{s\w}(t)\nonumber\\
		\frac {d \a_r(t)} {dt}=
		&-i\eps_r\a_r(t)-i\sqrt{\frac{\g}{2\pi}}\int^{\infty}_{-\infty}d\w \b_\w(t)e^{i\w \tau}-i\sqrt{\frac{\tw_s}{2\pi}}\int^{\infty}_{-\infty}d\w\c^\dagger_{r\w}(t)\label{aEQS}
\end{align}
Formally integrating~\eqref{bcEQS} for $\b_\w,\c_{r\w},\c_{s\w}$ and substituting in~\eqref{aEQS} 
we obtain
\begin{align}
\label{Langevin1}
	\frac{d\a_s(t)}{dt}=&-i\eps_s\a_s(t)
	-\sqrt{\g}\;\b_{in}(t)-\frac{\g}{2}\a_s(t)
	+\sqrt{\tw_s}\;\c^\dagger_{s}(t)+\frac{\tw_s}{2}\;\a_s(t) \nonumber\\
	\frac{d\a_r(t)}{dt}=&-i\eps_r\a_r(t)
	-\sqrt{\g}\;\b_{in}(t{-}\tau)
	-\frac{\g}{2}\a_s(t{-}\tau)
	-\frac{\g}{2}\a_r(t)
	+\sqrt{\tw_r}\;\c^\dagger_{r}(t)+\frac{\tw_r}{2}\;\a_r(t).
\end{align}
where we have defined the $\delta-$correlated {\it white noise operators} $\b_{\text{in}},\c_{k}$ 
\begin{align}
	\b_{\text{in}}(t)=\frac{i}{\sqrt{2\pi}}\intw e^{-i\w t}\b_\w(t) \;\;\;\;\text{and}\;\;\;\;
	\c_{k}(t)=\frac{i}{\sqrt{2\pi}}\intw e^{i\w t}\c_{k\w}(t)\label {noise}
\end{align}
which satisfy the commutation relations $[\b_{\text{in}}(t),\b_{\text{in}}^\dagger (t')]{=}\delta(t{-}t')$ and $[\c_{k}(t),\c_{k}^{\dagger }(t)]{=}\delta(t{-}t')$. In the following, with no loss of generality, we will neglect any field delay~\cite{gardiner1993} and we will take the limit $\tau\rightarrow0^+$ in~\eqref{Langevin1} to obtain
\begin{align}
\label{LangevinEQS}
	&\frac{d\a_s(t)}{dt}=(-i\eps_s+\frac{\tw_s}{2}-\frac{\g}{2})\;\a_s(t)
	-\sqrt{\g}\; \b_{\text{in}}(t)+\sqrt{\tw_s}\; \c^{\dagger}_{s}(t) \,, \nonumber \\
	&\frac{d\a_r(t)}{dt}=(-i\eps_r+\frac{\tw_r}{2}-\frac{\g}{2})\;\a_r(t)-\g\; \a_s(t)
	-\sqrt{\g}\; \b_{\text{in}}(t)+\sqrt{\tw_r}\; \c^{\dagger}_{r}(t) \,.
\end{align}
Note that while the input noise operator on $s$ is $\b_{\text{in}}(t)$, the input noise operator on $r$ is $\b_{\text{in}}(t)+\sqrt{\g} \, \a_s(t )$, which is the input-output relation due to a cascaded interaction with the travelling field modes~\cite{gardiner1993}. It implies that the dynamics of $r$ is driven by the output field from $s$ (but not the reverse).
The effective coupling between the two oscillators due to their cascaded interaction is more clearly visible when one describes their reduced dynamics in terms of the following {\it Master Equation}~\cite{giovannetti2012a,giovannetti2012,ramos2014,ramos2016} 
\begin{align}
	\frac{d\rh_{sr}(t)}{dt}=-i[\sum_{k=s,r}\eps_k\,\ac_k\a_k+H_{\text{casc}}^{sr},\rh_{rs}(t)]+
	\left(\g\mathcal{D}[\a_s{+}\a_r]+\tw_s\mathcal{D}[\ac_s]+\tw_r\mathcal{D}[\ac_r]\right)\rh_{sr}(t)
\label{ME}
\end{align}
where $\mathcal{D}[\op]\rh =2\op\rh\opc{-}\opc\op\rh{-}\rh\opc\op$.  In~\eqref{ME} one can identify the local pumps $\tw_s\mathcal{D}[\ac_s]$ and $\tw_r\mathcal{D}[\ac_r]$ while the effects of the dissipative cascaded coupling to the travelling modes is described jointly by a  collective decay $\g\mathcal{D}[\a_s{+}\a_r]$ and by the effective chiral hamiltonian, 
\begin{align}
	H_{\text{casc}}^{sr}=-i\frac{\g}{2}\left(\a_r\ac_s-\a_s\ac_r\right) \,.
\label{Hcasc}
\end{align}
proportional to $\g$ and hence of purely dissipative origin~\cite{lodahl2017}. Indeed the cascaded coupling of a pair of systems (HOs, two level systems etc.) with travelling modes leads to a collective dissipation of the pair of systems with a common bath and an effective interaction hamiltonian between the two, both effects being proportional, as pointed out, to the same constant $\g$. Dissipation and effective coupling therefore are not independent from each other. Their interplay plays a crucial role on the onset of synchronisation.


The two dynamical quantities which we will analyse to study the emergence of synchronisation are the mean value of the local oscillations and their two node correlations.
From~\eqref{H} and~\eqref{LangevinEQS} one can derive the following equation of motion for the average values $\av{\bm{\a}}{=}\lbrace\av{\a_1},\dots,\av{\a_N}\rbrace$ 
\begin{align}
 	\frac{d\av{\bm{\a}}}{dt}=\bm{M}\av{\bm{\a}}=\frac{1}{2}(\bm{W}{-}\bm{\Gamma}{-}2i\bm{\Omega}{-}2\g\bm{A})\av{\bm{\a}}\label{AvEqs}.
\end{align}
where $\bm{W}$ is the diagonal matrix of the local pump strength, $\bm{\Omega}$ is the diagonal matrix of the local energies and $\bm{\Gamma}$  is a diagonal matrix with entries
\begin{align}
 	\Gamma_{ii}=\g\sum_j(A_{ij}+A_{ji})\,, 
\end{align}
and where $\sum_j A_{ij}$ is the total number of edges from $i$ to any other node, while $\sum_j A_{ji}$ is the number of edges from any other node to $i$.
The off-diagonal entries of $\bf{M}$ correspond to the cascaded coupling between nosed and are all proportional to $\g$.

The correlations between amplitude fluctuations are conveniently characterized in terms of the covariance matrix  whose entries are  $C_{kj}{=}\frac{1}{2}\av{\chi_k\chi_j{+}\chi_j\chi_k}$, where
$\mathbf{ \chi}{=}\lbrace \a_1,\dots,\a_N,\ac_1,\dots,\ac_N\rbrace$~\cite{Olivares2012b}. 
Following a standard recipe~\cite{gardiner2004a}, we get a Lyapunov equation for the covariance matrix:

\begin{align}
	\frac{d\bm{C}}{dt}=
	\begin{pmatrix}\bm{M}&0\\0&\bm{M}^*\end{pmatrix}\bm{C}+
	\bm{C}\begin{pmatrix}\bm{M}&0\\0&\bm{M}^*\end{pmatrix}^T+
	\begin{pmatrix}0&\bm{S}\\\bm{S}&0\end{pmatrix}
\end{align}
with $\bm{S}=\frac{1}{2}[\bm{W}{+}\bm{\Gamma}{+}\g(\bm{A}{+}\bm{A}^T)]$.

Note that, both the time evolution of the annihilation (and creation) operators and of the correlation functions involve the matrix $\bm{M}$, although the relevant equations on motion are different~\cite{Roccati_2021}. 


\section{Synchronisation witnesses}\label{sec:synchronisation}

To characterise the emergence of synchronisation in our networks we will use two standard quantifiers~\cite{galve2017}. The first one, widely used in classical contexts to detect the synchronisation of two signals, $a(t)$ and $b(t)$, is the Pearson Factor~\cite{galve2017} defined as:
\begin{align}
	\Pearson{ab}(t|\Delta t)=\dfrac{\int_t^{t+\Delta t}\!\left(a(\tau){-}\overline{a}\right)\left(b(\tau){-}\overline{b}\right)d\tau}
	{\int_t^{t+\Delta t}\!\left(a(\tau){-}\overline{a}\right)^2\!d\tau\;\int_t^{t+\Delta t}\!\left(b(\tau){-}\overline{b}\right)^2\!d\tau}
	\label{Pearson1}
\end{align}
where the bar stands for a time average over the time window $\Delta t$.
\begin{align}
	\overline{a}=\dfrac{1}{\Delta t}\int_t^{t+\Delta t}a(\tau)d\tau
\end{align}
As a slight generalisation of ~\eqref{Pearson1}, we will consider also its phase shifted version, where the two time averages are evaluated over two shifted time windows, $\Delta t_a$ and $\Delta t_b$, in order to make them in phase. 
The Pearson Factor quantifies the correlation in the time domain between classical signals~\cite{pikovsky2001,boccaletti2002}. In the quantum domain the trajectories $a(t)$ and $b(t)$ can be the expectation values of any pair of quantum operators of interest~\cite{giorgi2012,manzano2013,giorgi2019} like $\av{a}$,$\av{x}$,$\av{x^2}$ etc.
Furthermore to detect the onset of synchronisation between more that two nodes in a sub-network $\bm{g}$,  we simply take the product of the Pearson factor for pairs of nodes of the subnetwork of interest~\cite{cabot2018}:
\begin{align}
	S_{\bm{g}}(t|\Delta t)=\prod_{ij\in {\bm{g}}}\Pearson{a_ia_j}(t|\Delta t)
	\label{Sind}
\end{align}
A second way to characterise the emergence of synchronisation, particularly suited for periodic evolution, is to evaluate the dynamical Fourier transform of the signals of interest in the same time window used to evaluate the Pearson factor, more precisely:
\begin{align}
	f_a(\eps,t|\Delta t)=\frac{1}{\Delta t}\int_t^{t+\Delta t}  e^{-i\eps\tau}a(\tau)d\tau
\label{DF}
\end{align}
In this case  the emergence of synchronisation manifest itself in the evolution of a spectrum initially involving several frequencies towards a single-frequency spectrum.

In the following the signals which will enter in~\eqref{Pearson1} or~\eqref{DF} will be either the complex amplitudes $\av{\hat{a}_k}$ and the symmetrized second order moments entries of the covariance matrix of the local oscillators' quadrature 
$\av{\hat{x}_k^2}$ and  $\av{\hat{x}_k\hat{y}_k+\hat{y}_k\hat{x}_k}$ where $\hat{x}_k = (\hat{a}_k + \hat{a}^{\dagger}_k)/2$ and $\hat{y}_k = - i(\hat{a}_k - \hat{a}^{\dagger}_k)/2$

The synchronisation of the quadrature fluctuations can be a signature of non local correlations in the dynamics of the pair of HOs~\cite{Galve2012}. To assess the quantum nature of such correlation between the pair of HOs $i$ and $j$  we will use the so called Quantum Discord~\cite{ollivier2001,modi2012,bera2017}, defined as 
\begin{equation}
{\mathcal D}_{ij} = S (\rho_i) - S(\rho_{ij} + \min_{{\hat \pi}_k}\sum_k p_k S(\rho_{i|k})
\end{equation}
with $S(\rho ) = -\text{tr} (\rho\ln\rho)$ the von Neumann entropy and the minimisation is taken over all possible quantum measurement $\{{\hat \pi}_k\}$ made on HO $j$. 
A measurement outcome with index $k$ will induce a collapse of the joint density operator $\rho_{ij}$ into the reduced operator $\rho_{i|k} = \text{tr}_j (\pi_k\rho_{ij})/p_k$ of the HO $i$. A non zero value of the discord is the signature of a quantum nature of the correlations between $i$ and $j$. Note that in general (and in our particular setup) the discord is asymmetric i.e. ${\mathcal D}_{ij} \neq {\mathcal D}_{ji}$ and overlying arrows will be used in the following to distinguish them. For Gaussian states the minimisation can be be restricted to gaussian measurements~\cite{giorda2010,adesso2010}. In the following we therefore use as quantifier the gaussian discord for which there is a closed albeit cumbersome analytical form~\cite{adesso2010}.  
\section{Simple Graphs}\label{sec:simplegraph}
As we have anticipated the kind of synchronisation pattern depends strongly on the network structure. To gain an insight into such possible  synchronisation patterns we will analyse some illustrative example of growing complexity.

\paragraph{Linear chains --- } Linear chains are just linear graphs with pairwise edges (couplings). As we will show such structures can exhibit both  synchronisation or frustration depending on the directionality of the couplings.\\

\begin{tikzpicture}[scale=1,baseline=-.5ex]
	\node (1) at (0,0) [draw, circle,fill=black,inner sep=0pt,minimum size=7pt]{};
	\node (2) at (0.5,-0.) [draw, circle,fill=black,inner sep=0pt,minimum size=7pt]{};
	\draw[->,line width=1.5pt]	(1) -- node {} (2);
\end{tikzpicture}
{\bf Oriented Dimers}, i.e.  pairs of cascaded coupled oscillators, are the simplest structure exhibiting  synchronisation. They are also the only structure in which the whole network interacts with a single common environment in our model. The matrix ${\bf M}$ in this case takes the form:
\begin{align}\label{eq:MatrixM_simnplest}
	\bm{M}=\frac{1}{2}\begin{pmatrix}-2i\eps_1+\tw_1-\g&0\\-2\g&-2i\eps_2+\tw_2-\g\end{pmatrix} \,.
\end{align}
and the  mean values   $\av{\bm{a}}$ evolve as:
\begin{align}
	\av{a_1(t)}&=e^{-\frac{t}{2}(\g{-}\tw_1{+}2i\eps_1)}\av{a_1(0)}\label{pair_sol}\\
	\av{a_2(t)}&=e^{-\frac{t}{2}(\g{-}\tw_2{+}2i\eps_2)}\av{a_2(0)}+
	2\g	\frac{(e^{-\frac{t}{2}(\g-\tw_2{+}2i\eps_2)}{-}e^{-\frac{t}{2}(\g-\tw_1{+}2i\eps_1)})}{(\tw_1{-}\tw_2){-}2i(\eps_1{-}\eps_2)}\av{a_1(0)}\nonumber
\end{align}
From~\eqref{pair_sol} we see that for  $\g{-}\tw_1\sim0$ and $\g{-}\tw_2>0$, after a transient, 
we observe synchronisation, as all the terms oscillating at frequency $\eps_2$ vanish. Synchronization will be stationary~\cite{cabot2018} when pumping compensate losses in the origin node 1, i.e. for $\g \equiv \tw_1$)
\begin{figure}[!ht]
	\includegraphics[width=0.9\linewidth]{ 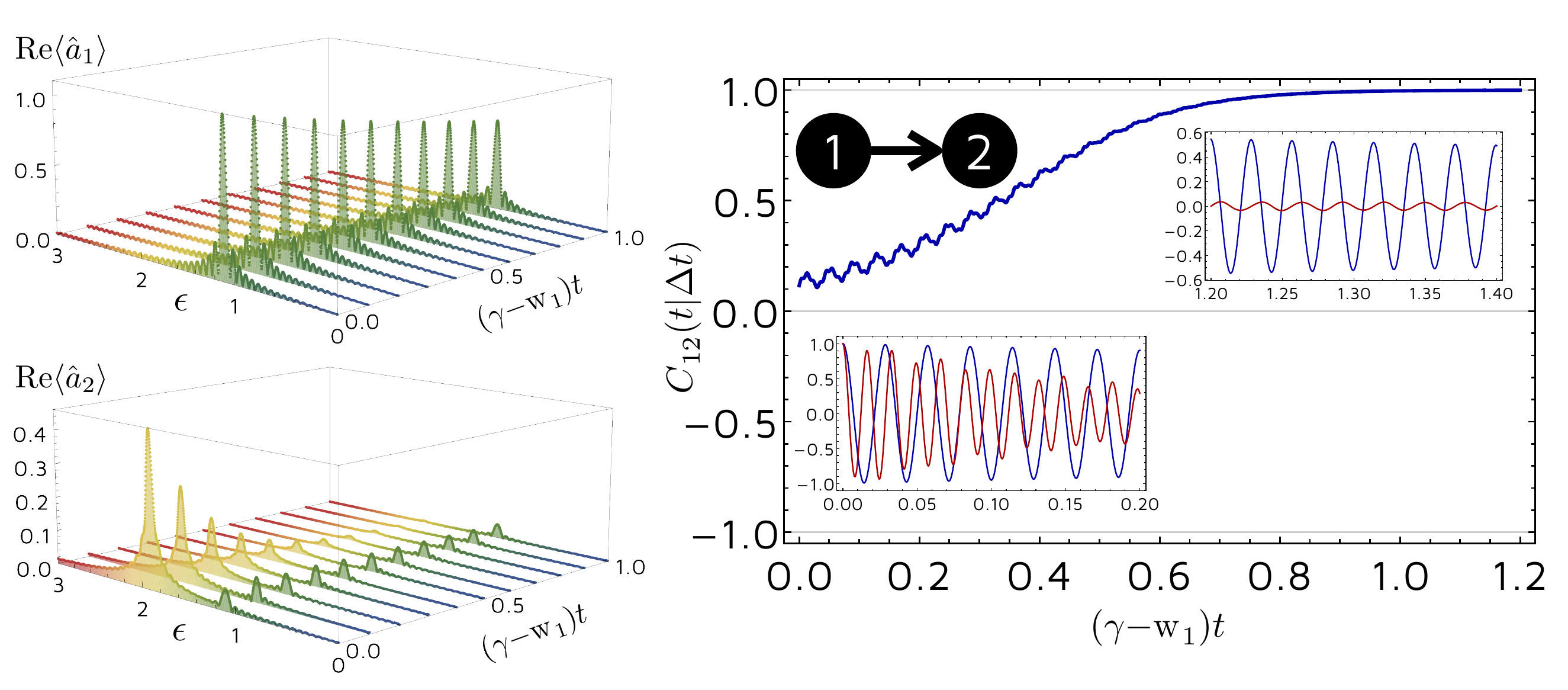}
	\caption{Synchronisation of a dimer dynamics with $\g=0.05$, $\tw_1=0.9\,\g$, $\tw_2=0$. The local frequencies are $\eps_1=1,\eps_2=1.9$. In all the plots the time is taken in $\g-\tw_1$ units. (Left) Dynamical Fourier transform of $\text{Re}\av{\a_1}$ (left-top) and $\text{Re}\av{\a_2}$ (left-bottom). (Right) The Pearson Factor~\eqref{Pearson1} associated with $\text{Re}\av{\a_1}$ and $\text{Re}\av{\a_2}$. In the insets are plotted the time evolutions of $\text{Re}\av{\a_1}$ (blue line) and $\text{Re}\av{\a_2}$ (red line) in different time windows.}
	\label{fig:fig2}
\end{figure}
Note  furthermore that the greater the difference $\tw_1-\tw_2$, the lower the amplitude of the residual signal $2$, while  a smaller difference leads to a persistence of the component at  frequency $\eps_2$ in the second signal.  On the other hand, for larger values of $\g$ one observes a faster decay.
In \figref{fig:fig2} we set  $\g=0.05$, $\tw_1=0.9\g$ and $\tw_2=0$. 
In \figref{fig:fig2}a, we plot the dynamical Fourier transform of $\text{Re}\av{\a_1}$ and $\text{Re}\av{\a_2}$. It is evident that the two oscillators after a transient synchronise as witnessed by the Pearson factor of the two signals  $\Pearson{a_1a_2}(t|\Delta t)$ plotted in \figref{fig:fig2}b. 
If pumping is removed no synchronization occurs between detuned oscillators, being this a major difference of the chiral case with respect to the undirected one~\cite{Galve2012}.

\begin{figure}[!ht]
	\centering
	\includegraphics[width=0.9\linewidth]{ 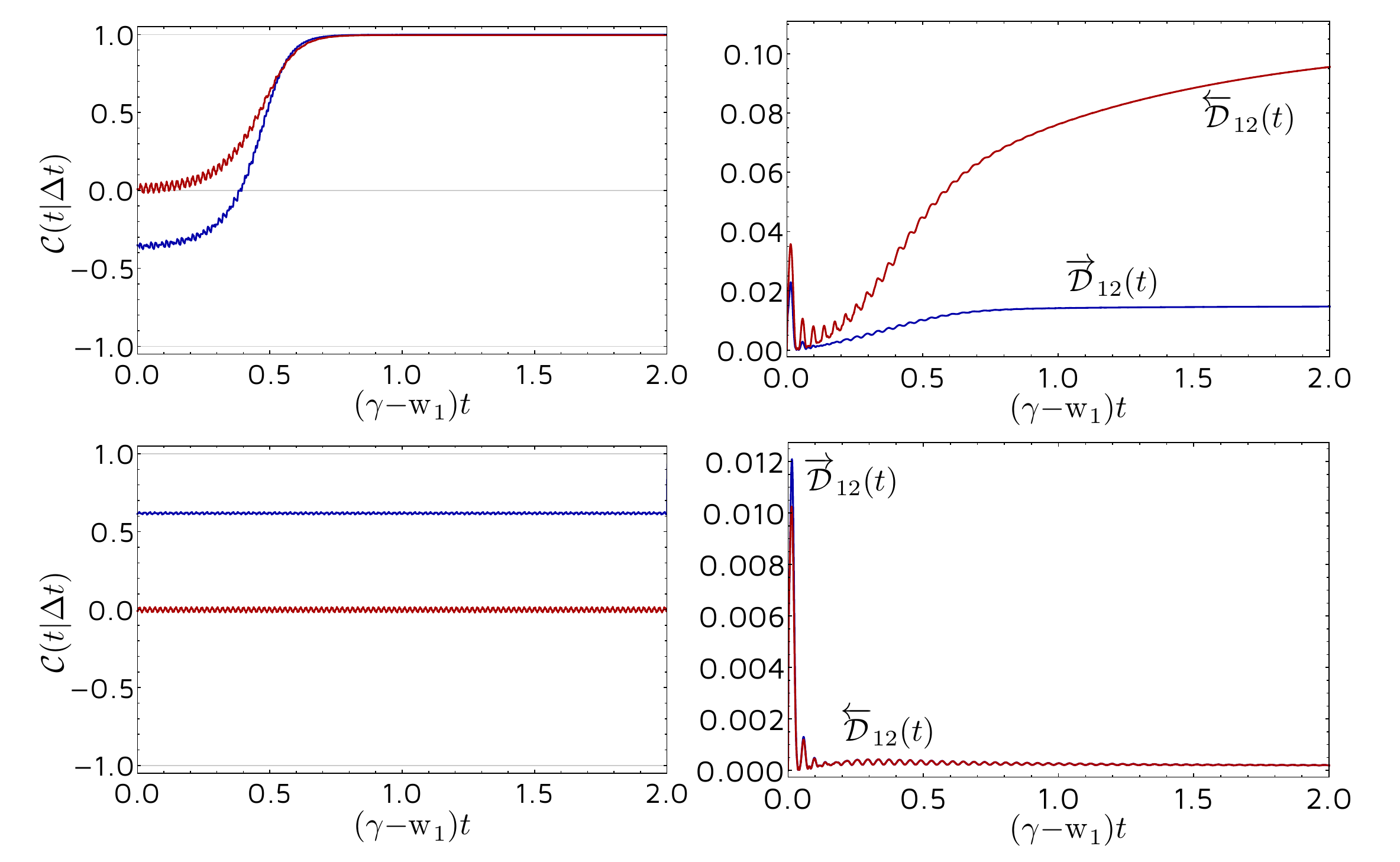}
	\caption{(Top): Synchronisation of correlations in a cascaded dimer, same parameters as in \figref{fig:fig2}. (Left) Pearson factor relative to $\{ \av{\hat{x}_1^2},\av{\hat{x}_2^2}\}$ (blue curve) and  $\{\av{\hat{x}_1\hat{y}_1+\hat{y}_1\hat{x}_1},\av{\hat{x}_2\hat{y}_2+\hat{y}_2\hat{x}_2}\}$ (red curve). (Right) Quantum Discord in the two directions: $1\rightarrow2$ (blue curve) and $1\leftarrow2$ (red curve). (Bottom): Same parameters as before except for the pump of the second oscillator that in this case is $\tw_2=\tw_1=0.9\g$. As clarified in the main text, equally pumping the two nodes causes loss of synchronisation and at the same time also quantum correlations vanish. }
	\label{fig:fig3}
\end{figure}

The onset of synchronisation not only of the average values but also of the local the local quadratures is shown in \figref{fig:fig3} where we have plotted the Pearson factor for the pair of signals  $\{ \av{\hat{x}_1^2},\av{\hat{x}_2^2}\}$ and symmetrised covariance elements $\{\av{\hat{x}_1\hat{y}_1+\hat{y}_1\hat{x}_1},\av{\hat{x}_2\hat{y}_2+\hat{y}_2\hat{x}_2}\}$. To elucidate the quantum nature of the mutual correlations of the synchronised motion we have evaluated the quantum discord between two HOs and plotted it in \figref{fig:fig3}, where it is shown that quantum correlations are present in the time window where the system exhibits synchronisation.

\begin{figure}[!ht]
	\centering
	\includegraphics[width=0.9\linewidth]{ 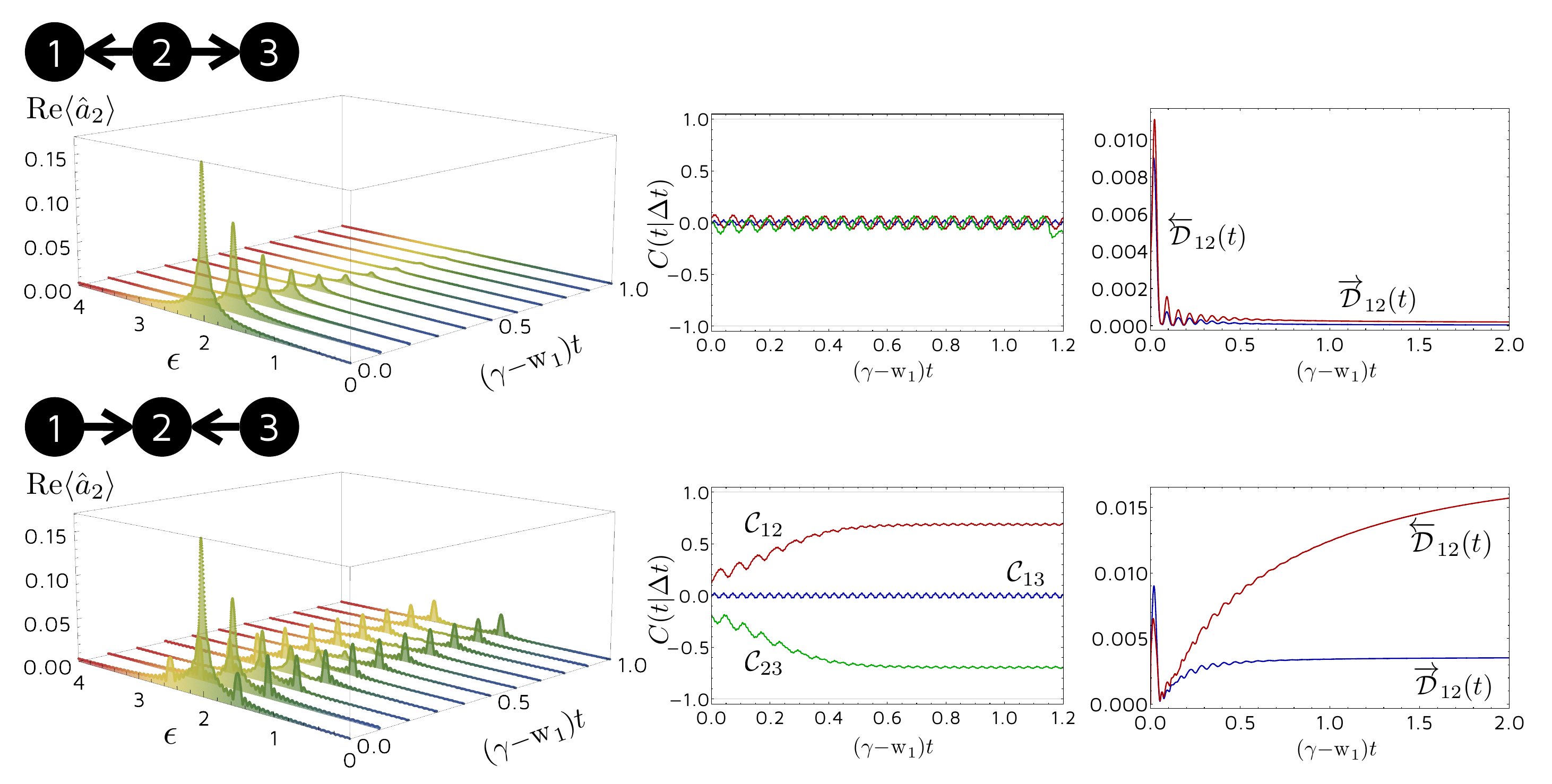}
	\caption{Synchronisation of linear three nodes configurations in which we set the frequencies of HOs $\epsilon_n{=}\lbrace{1.5,,2,2.5\rbrace}$, the decaying rate $\gamma=0.05$ and the pumping intensity $\tw=0.045$. Dynamical Fourier Transform (left), Pearson factors for the average motion (center), discord (right)}
	\label{fig:fig4}
\end{figure}

\begin{figure}[!ht]
	\centering
	\includegraphics[width=0.9\linewidth]{ 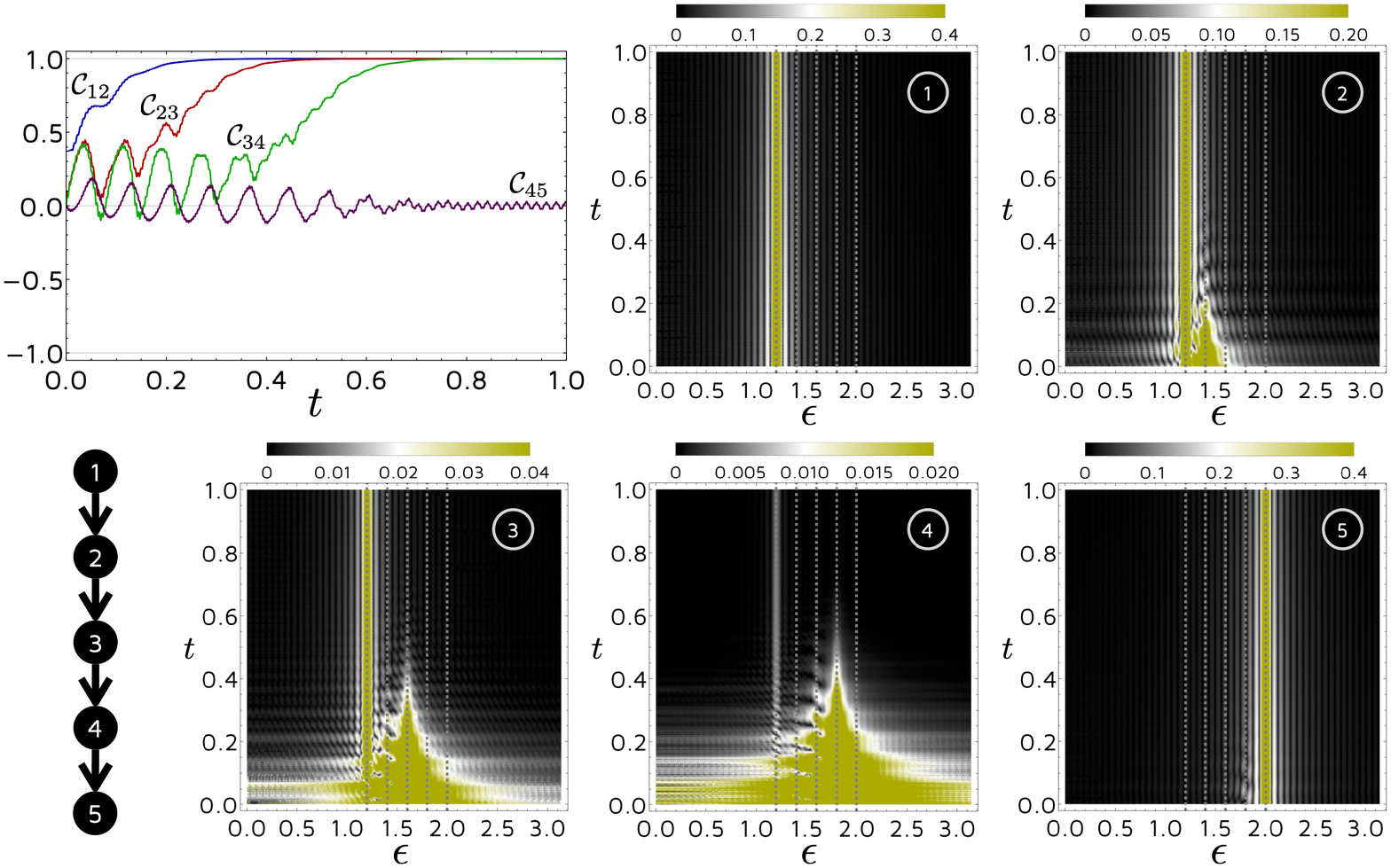}
	\caption{Synchronisation of a linear chain with five nodes. The local frequencies of the HOs are  $\epsilon_n{=}\lbrace{1.2,1.4,1.6,1.8,2\rbrace}$, the decaying rate $\gamma=0.05$ and the pump intensity $w_n=0.045$. The time is in unit $\/(\g-\tw)$.
	The top left figure shows the shifted Pearson factor $C_{kj}(\Delta t)$ (optimized with delay) relative to different pairs of sites.
	For completeness we report also the five spectra of the nodes evaluated in the same time window $\Delta t$ used for evaluation of the Pearson factor. It is evident how all the nodes, but $5$, synchronise with the first frequency at different times.}\label{fig:fig5}
\end{figure}

As we are going to analyse more complex networks we will assume, from now on,  equal local pump intensity for each  harmonic oscillator. As we have seen for dimers, a suitable engineering of the pattern of pump strengths leads always to synchronisation. With uniform pump strengths instead, neither the dimers nor some of the three-node configurations
we will characterise in the following, synchronise.\\
We proceed our analysis of small network by looking at linear chains 
 with three nodes (see \figref{fig:fig4} and \figref{fig:fig5}). For undirected networks both transient and stationary synchronization were reported in presence of a common bath tuning frequencies and couplings~\cite{manzano2013a}. In the following  directional couplings determine different behaviours, as also the presence of pumping.\\
 
\begin{tikzpicture}[scale=1,baseline=-.5ex]
	\node (1) at (0,0) [draw, circle,fill=black,inner sep=0pt,minimum size=7pt]{};
	\node (2) at (0.5,0.) [draw, circle,fill=black,inner sep=0pt,minimum size=7pt]{};
	\node (3) at (1.,0) [draw, circle,fill=black,inner sep=0pt,minimum size=7pt]{};
	\draw[->,line width=1.5pt]	(2) -- node {} (1);
	\draw[->,line width=1.5pt]	(2) -- node {} (3);
\end{tikzpicture}
{\bf Trimer out} In this first scenario, the central node drives the two external ones, each with its own local frequency. For this configuration the equations for the average values have the following solution:
 \begin{align}
	\av{a_1(t)}&=e^{-\frac{t}{2}(\g{-}\tw{+}2i\eps_1)}\av{a_1(0)}+2\g
	\frac{(e^{-\frac{t}{2}(2\g-\tw{+}2i\eps_2)}{-}e^{-\frac{t}{2}(\g-\tw{+}2i\eps_1)})}{\g{-}2i(\eps_1{-}\eps_2)}\av{a_2(0)}\nonumber\\
	\av{a_2(t)}&=e^{-\frac{t}{2}(2\g{-}\tw{+}2i\eps_2)}\av{a_2(0)}\nonumber\\
	\av{a_3(t)}&=e^{-\frac{t}{2}(\g{-}\tw{+}2i\eps_3)}\av{a_3(0)}+2\g
	\frac{(e^{-\frac{t}{2}(2\g-\tw{+}2i\eps_2)}{-}e^{-\frac{t}{2}(\g-\tw{+}2i\eps_3)})}{\g{+}2i(\eps_2{-}\eps_3)}\av{a_2(0)}\nonumber
 \end{align}
Surprisingly, the central node does drive but does not synchronize the edges. In fact the dynamics at frequency $\eps_2$ decays at a larger rate $2\g$ due to the fact that node $2$ has a double decay channel. As a consequence, driving is inhibited and for $\tw{-}\g{\sim}0^-$, at long time, the first and third signals oscillate at their own frequencies, showing no synchronisation independently on the central oscillator frequency) (see \figref{fig:fig4}).
Also larger pumping will not help, as the edge oscillations (at the local frequencies) will always dominate. Again if we admits different pumping intensities, more precisely $\tw_2>\tw_{1,3}$, the central node is able to impose its frequency to the others, reaching a global synchronized motion.\\

\begin{tikzpicture}[scale=1,baseline=-.5ex]
	\node (1) at (0,0.) [draw, circle,fill=black,inner sep=0pt,minimum size=7pt]{};
	\node (2) at (0.5,0) [draw, circle,fill=black,inner sep=0pt,minimum size=7pt]{};
	\node (3) at (1.,0.) [draw, circle,fill=black,inner sep=0pt,minimum size=7pt]{};
	\draw[->,line width=1.5pt]	(1) -- node {} (2);
	\draw[->,line width=1.5pt]	(3) -- node {} (2);
\end{tikzpicture} {\bf Trimer in} When we revert both edge directions the two external nodes drive the central one. In this geometry the central node undergoes a frustrated dynamics as we can see from the following solution:
\begin{align}
	\av{a_2(t)}=&e^{-\frac{t}{2}(2\g{-}\tw{+}2i\eps_2)}\av{a_2(0)}+\nonumber\\
	&2\g\frac{(e^{-\frac{t}{2}(2\g-\tw{+}2i\eps_2)}{-}e^{-\frac{t}{2}(\g-\tw{+}2i\eps_1)})}{\g{+}2i(\eps_2{-}\eps_1)}\av{a_1(0)}+
	2\g\frac{(e^{-\frac{t}{2}(2\g-\tw{+}2i\eps_2)}{-}e^{-\frac{t}{2}(\g-\tw{+}2i\eps_3)})}{\g{+}2i(\eps_2{-}\eps_3)}\av{a_3(0)}\nonumber \,,
\end{align} 
while each of the other two nodes evolves oscillating at its own frequency and decays at its relevant rate.
Once again the central node has two edges and so undergoes a double decay, but in this case its evolution is driven by both $1$ and $3$.
In \figref{fig:fig4} we plot the dynamical Fourier transform~\eqref{DF} of the central node average amplitude. After a transient, we observe a two peak spectrum, at frequencies $\epsilon_1$ and $\epsilon_3$. Note that the amplitude of the signal is rather small, compared for example to that of \figref{fig:fig2}. This is due to the fact that in this case we have equal pump strengths for the three oscillators, at variance with the two-node case previously analysed.  \\

\begin{tikzpicture}[scale=1,baseline=-.5ex]
	\node (1) at (0,0) [draw, circle,fill=black,inner sep=0pt,minimum size=7pt]{};
	\node (2) at (0.5,0.) [draw, circle,fill=black,inner sep=0pt,minimum size=7pt]{};
	\node (3) at (1.,0) [draw, circle,fill=black,inner sep=0pt,minimum size=7pt]{};
	\draw[->,line width=1.5pt]	(1) -- node {} (2);
	\draw[->,line width=1.5pt]	(2) -- node {} (3);
\end{tikzpicture} {\bf Trimer through} While in the previous cases we have no synchronisation, reversing the direction of only one link (leaving all the parameters unchanged), we have a partial synchronisation. 
This fact shows the major difference between an undirected and a directed network. 
Generalizing this setup to chain of $N$ HOs, in fact (see \figref{fig:fig5} for the $5$ HOs case) we have that all the nodes, but the last, synchronize at frequency of the first one. The first HO is able to force synchronisation on the rest of the chain but not on the last because the $N$-th HO has the same degree of the first one and hence the pumping turns out to be too intense. 
In \figref{fig:fig5} is reported the synchronisation quantifier $\Pearson{{\av{a_i}\av{a_j}}}$ for a concatenation of $N=5$ HOs, it is evident how after a transient all following nodes synchronize one at a time with the first, except the last one. Indeed, each HO, once it has synchronized with the previous ones, forces the next HO to synchronize. This implies a sort of propagation of the synchronisation phenomenon from the first HO of the chain towards the last one. As discussed before, the last one does not synchronize.
\paragraph{Rings and branches --- } Let us now consider some examples of non linear graphs introducing rings or bifurcations in our networks.\\

\begin{tikzpicture}[scale=1,baseline=-.5ex]
	\node (1) at (0,0) [draw, circle,fill=black,inner sep=0pt,minimum size=7pt]{};
	\node (2) at (0.3,0.4) [draw, circle,fill=black,inner sep=0pt,minimum size=7pt]{};
	\node (3) at (0.6,0) [draw, circle,fill=black,inner sep=0pt,minimum size=7pt]{};
	\draw[->,line width=1.5pt]	(1) -- node {} (2);
	\draw[->,line width=1.5pt]	(2) -- node {} (3);
	\draw[->,line width=1.5pt]	(1) -- node {} (3);
\end{tikzpicture} {\bf  Non-loop ring}
The simplest ring geometry is a three nodes network in which each pair of nodes is coupled via  common travelling modes. Let us first consider the case in which node 1 drives nodes 2 and 3 while node 3 is driven by nodes 1 and 2. In this case we have 
\begin{equation}\label{MMatrixN}
	\bm{M} = \frac{1}{2}
	\begin{pmatrix}
		\eta_1 & 0 & 0 \\
		-2\g & \eta_2 & 0 \\
		-2\g & -2\g & \eta_3 \\
	\end{pmatrix},,
\end{equation}
where $\eta_j = - 2 i \epsilon_j - k\gamma + \tw $. 
Since the matrix is triangular, its eigenvalues are  equal to the diagonal elements: $\lambda_j {=} \eta_j$. Also in this case, for uniform local pumps,  nodes do not to synchronise. But If we choose all the pump rates for $j>1$ smaller than $\gamma$ ($\tw_j{-}\gamma < 0$), and $\tw_1{-}\gamma\sim 0$, we obtain that only the single collective mode associated to $\eta_1$ survives. This implies that after a transient all the local oscillators will oscillate at frequency $\eps_1$.\\

\begin{tikzpicture}[scale=1,baseline=-.5ex]
	\node (1) at (0,0) [draw, circle,fill=black,inner sep=0pt,minimum size=7pt]{};
	\node (2) at (0.3,0.4) [draw, circle,fill=black,inner sep=0pt,minimum size=7pt]{};
	\node (3) at (0.6,0) [draw, circle,fill=black,inner sep=0pt,minimum size=7pt]{};
	\draw[->,line width=1.5pt]	(1) -- node {} (2);
	\draw[->,line width=1.5pt]	(2) -- node {} (3);
	\draw[->,line width=1.5pt]	(3) -- node {} (1);
\end{tikzpicture} 
{\bf  Loop ring}
For a three nodes cyclic network in which each node drives the following one and is driven by the previous one 
the matrix $\bm{M}$ assumes the following form:
\begin{equation}
\bm{M} =\frac{1}{2} \left(
\begin{array}{ccc}
	\eta_1 &0 &-2\g \\
	-2\g & \eta_2 & 0 \\
	0 & -2\g& \eta_3 
\end{array}
\right)\,,
\end{equation}
with $\eta_j=-2i\epsilon_j -2\g+ \tw$. Also in this case we do not observe synchronisation, unless  the pump intensities are tuned to proper different values. \\

\begin{figure}[h!]
	\centering	
	\includegraphics[width=1\linewidth]{ 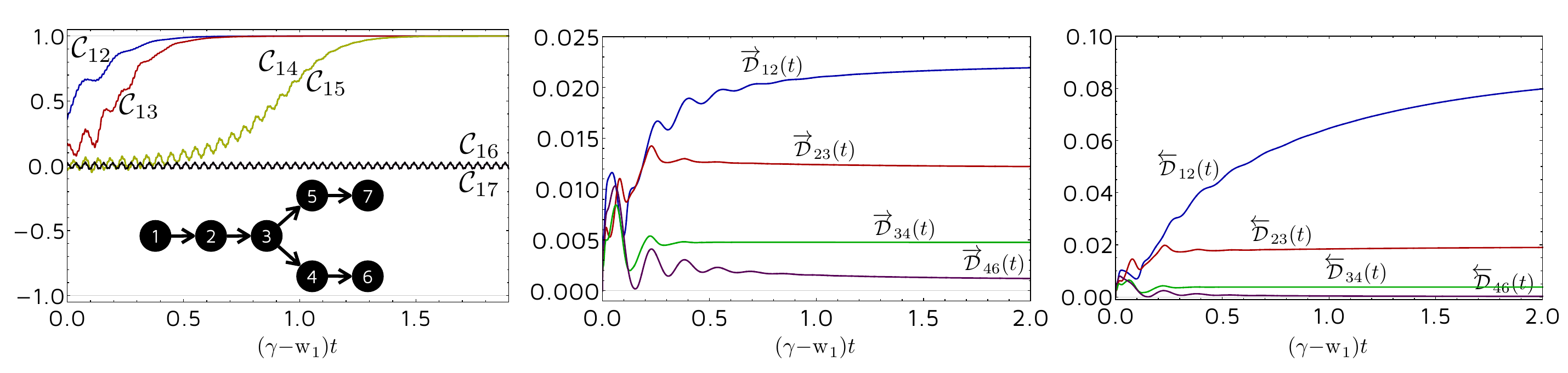}
	\caption{Synchronization of a seven nodes array with a bifurcation and corresponding Pearson factors (left pane). Discord (central and right pane). Parameters are set as in \figref{fig:fig5} with  equally spaced frequencies from $1.2$ to $2.2$ with nodes $4,5$ and $6,7$ having the same frequencies. The plot clearly show the lack of synchronization on the final nod of the two branches.} \label{fig:fig6}
\end{figure}

\begin{tikzpicture}[scale=1,baseline=-.5ex]
	\node (1) at (0,0) [draw, circle,fill=black,inner sep=0pt,minimum size=7pt]{};
	\node (2) at (0.5,0.0) [draw, circle,fill=black,inner sep=0pt,minimum size=7pt]{};
	\node (3) at (1,-0.3) [draw, circle,fill=black,inner sep=0pt,minimum size=7pt]{};
	\node (4) at (1,0.3) [draw, circle,fill=black,inner sep=0pt,minimum size=7pt]{};
	\draw[->,line width=1.3pt]	(1) -- node {} (2);
	\draw[->,line width=1.3pt]	(2) -- node {} (3);
	\draw[->,line width=1.3pt]	(2) -- node {} (4);
\end{tikzpicture} 
{\bf Branching chain} As a final simple non linear topology we now analyse branches. The simplest branch geometry consists of  four nodes: a first HO which drives a second, which in turn drives, independently, two further HOs. The relevant matrix $\bm{M}$ is:
\begin{equation}
\bm{M} =\frac{1}{2} \left(
\begin{array}{cccc}
	\eta_1 &0 & 0 & 0 \\
	-2\g & \eta_2 & 0 & 0\\
	0 & -2\g& \eta_3 & 0 \\
	0 & -2\g& 0 & \eta_4 \\
\end{array}
\right)\,.
\end{equation}
When allowing for longer chains after the branching, all HOs synchronize to the frequency of the first one, except for the last ones of each branch, as shown in \figref{fig:fig6} where the synchronization quantifier are plotted for a seven nodes branch. The reason for such lack of synchronisation of the last node of each branch is the same as discussed for the linear chain. Also in \figref{fig:fig6}, we see that synchronization among distant elements requires more time to be established (the left panel shows synchronization with the first node). As for the linear chain (not shown), the discord between pair of nodes decays along each chain and is not symmetric, displaying larger values  when measuring the input node.

\paragraph{Complex Networks --- } We finally consider a more complex scenario enclosing all the motifs seen so far.
Given a directed graph $\bm{G}$ with $N$ nodes and $K$ edges, described by the adjacency matrix $\bm{A}$, 
we associate to each node a frequency $\epsilon_i$ uniformly distributed in the interval $[\epsilon_{\text{min}},\epsilon_{\text{max}}]$ with the only constraint $|\epsilon_i -\epsilon_j| > \g$ for any pair of nodes $i,j$. In this scenario the eigenfrequencies of the matrix $\bm{M}$ turn out to be very close to the natural frequencies of the HOs.
The example shown in \figref{fig:fig5} represent a directed graph composed by $15$ HOs. In this configuration the network shows three  communities of synchronisation. After a transient time nodes $12$, $13$ and $14$ impose their frequencies to respective groups. Note that node $1$, that has three outgoing edges, does not synchronise with any other nodes.  Node $6$, driven by nodes $13$ and $10$, undergoes a frustrated dynamics as in the example \figref{fig:fig2}, but at the end synchronizes with $10$, given that  $10$ survives longer than 13 thanks to input from node $7$. In \figref{fig:fig5} (left) is reported the collective quantifier~\eqref{Sind} relative to the three communities.
\begin{figure}
	\centering
	\includegraphics[width=0.9\linewidth]{ 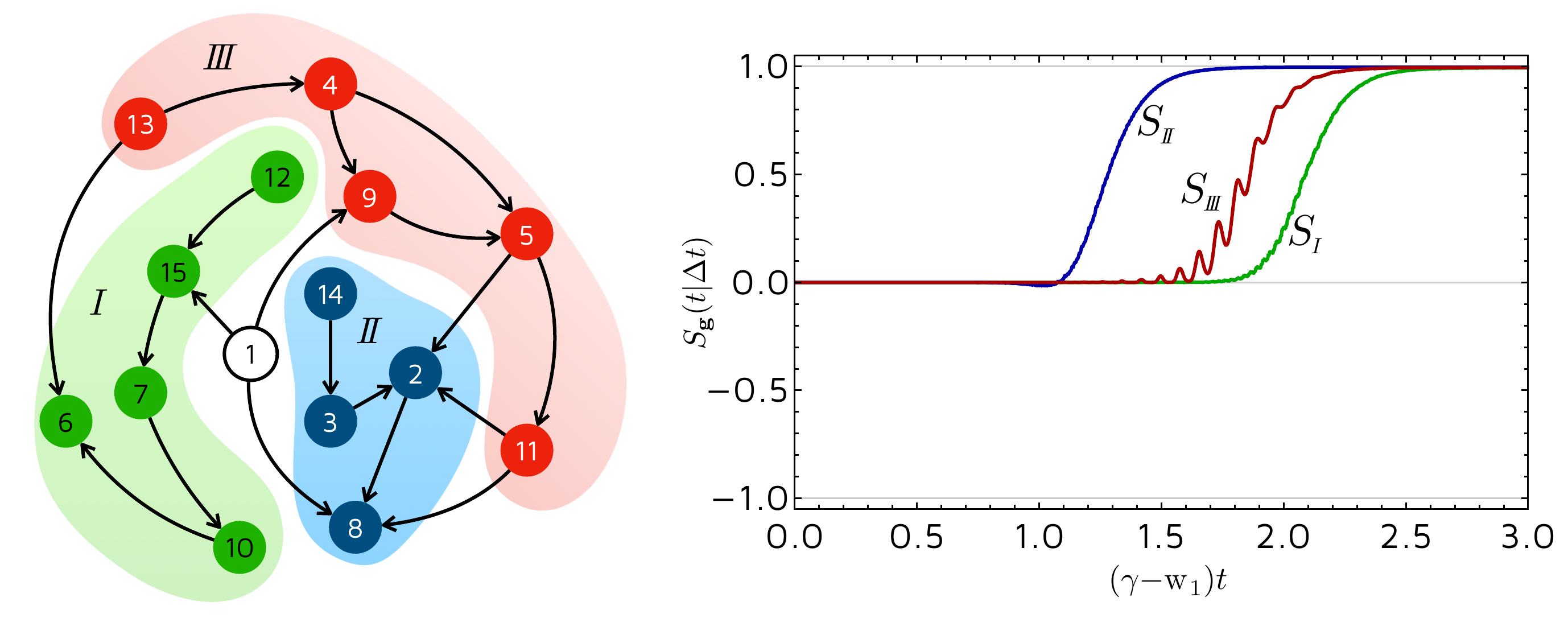}
	\caption{Example of a complex directed network. The frequencies of the harmonic oscillators are equally spaced in the interval $\{1.2 - 4\}$ (left). On the right it is reported the collective Pearson factor $	S_{\bm{g}}(t|\Delta t)$~\eqref{Sind} for the three different communities of synchronisation highlighted in the sketch to the left.}
	\label{fig:fig7}
\end{figure}

The spectrum of the matrix $\bm{M}$~\eqref{AvEqs} provides a clear insight into the emerging synchronisation patterns analysed so far, allowing to establish the presence of a time scale separation in the modes decays needed for synchronization~\cite{transientbookchapter}. 
For example in the simple case of dimers, the eigenvalues of the matrix $\bm{M}$ of~\eqref{eq:MatrixM_simnplest} are trivially the diagonal matrix elements, $\tw_1-\g$ and $\tw_2-\g$, and the corresponding (non-normalized) eigenvectors are $(-2i(\epsilon_2-\epsilon_1) + \tw_2-\tw_1, 2\g)$ and $(1, 0)$. Under the assumption $\tw_1 - g \sim 0$ and $\tw_2 - \g < 0$, one of the eigenvalues has its real part close to zero, corresponding to a long living mode, while the other mode decays faster. Since the surviving mode involves both  HOs, the long-time dynamics is obviously a synchronized motion.
In the ring topology corresponding to the matrix $\bm{M}$ of~\eqref{MMatrixN}, no synchronisation is possible for  equal local pumps, as in this case the three eigenvalues have equal real parts, which in turn implies that none of the three eigenstates survives to the other two.

\begin{figure}
	\centering
	\includegraphics[width=1\linewidth]{ 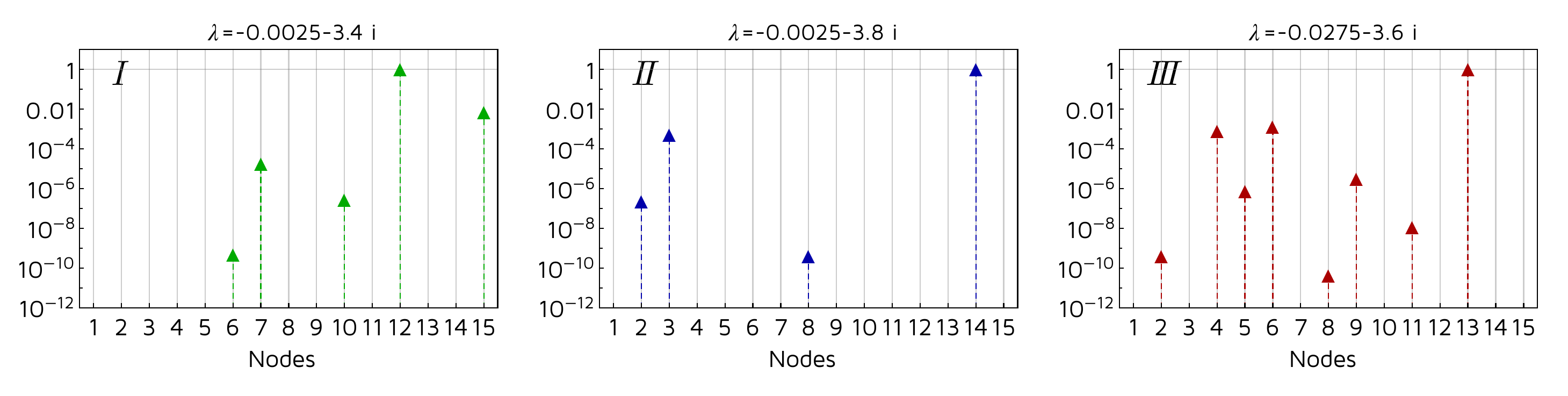}
	\caption{The squared absolute values of components of the three eigenstates responsible for the appearance of synchronisation clusters. In the top of each plot the relevant eigenvalue is reported.}
	\label{fig:fig8}
\end{figure}

Let us now focus on the complex network previously considered. In such a case, out of the fifteen eigenvalues (numerically evaluated, all with non positive real parts) we can identify the three of them with real parts closer to zero, i.e., the three which survive longer. Looking at the table plots in \figref{fig:fig8}, we see that the corresponding eigenstates roughly involve the nodes of the three clusters of \figref{fig:fig7}, with some exceptions which can be easily understood. The two eigenvectors corresponding to the two dominant eigenvalues (those with real parts equal to $-0.0025$) involve the nodes $(2,3,8,14)$ in II and $(6,7,10,12,15)$  in I , which perfectly fit two of the three clusters of synchronisation. The third eigenvalue (with real part $-0.0275$) involves all the following nodes: $(2,4,5,6,8,9,11,13)$. Still such cluster does not include nodes $2$ and $6$, whose long time dynamics is determined by the other two eigenstates, so that, after a sufficiently long time the oscillations of such two nodes are governed by the frequencies of the other two clusters. 
The presence of the isolated node $1$ is well explained looking at another eigenvector which involves node $1$ and other nodes already involved in the previously considered clusters .

\section{Discussion and conclusions}\label{sec:conclusion}

We have analysed the emergence of synchronisation in quantum chiral networks of HOs. Each HO is locally incoherently pumped and pairs of HOs interact with each other via a cascaded coupling with travelling modes. Such coupling gives rise to an effective pairwise interaction and to a pairwise collective loss, the two effects being of purely dissipative origin and dependent on the same dissipation parameter. Local incoherent coupling is also introduced allowing the system to display sustained oscillations. For symmetric bidirectional couplings reported so far, different forms of dissipation are known to induce synchronization in different systems, as reviewed in Ref.~\cite{transientbookchapter,cabot2019}.

A specific feature of directional couplings considered here is the possibility to induce hybrid  features of both mutual and driven synchronization arise. Indeed the directional coupling establishes a different (driver or slave) role between nodes.

We have analysed networks of increasingly complex topologies. Even in the simplest configuration, as a dimer, a chiral coupling changes the synchronization scenario. Indeed collective dissipation for bidirectional coupling induces mutual synchronization~\cite{Galve2012} but in the chiral configuration synchronization cannot be established unless pumping is included.
As we have shown for the cascaded dimer, where the first HO drives without feedback the second one, the appropriate choice of the local pumping gives rise to the onset of pairwise synchronisation. Furthermore, for equal local pumping, no synchronisation occurs.  If longer chains are considered with couplings all in the same directions, we find long-distance synchronization in agreement with Ref.
\cite{li2016}, where unidirectionally cascaded optomechanical systems have been considered in the classical regime.

We have then extended our analysis to more complex geometries, from unidimensional graphs to rings and branches. For non-chiral couplings global dissipation has been shown to induce transient synchronization in small (breathers) motifs in different complex networks in Ref.~\cite{cabot2018}. Within this chiral quantum optics set-up the scenario changes both for the couplings directionality and for the form of dissipation. In all cases, we have seen that, with fixed parameters (local energies and pumping), {\it local} topological changes (for example the addition of a link or even the simple swap of direction of a link) have a strong influence on the onset of synchronisation, meaning that in the general situation sub-networks synchronise independently and synchronised clusters emerge. 
In most of the situations, an analysis of the quantum discord shows that the establishment of a synchronised motion of two oscillators is associated with an increase of mutual quantum correlations. 

The analysis of the eigenvalues of the matrix $\bm{M}$ makes it possible to predict the onset of synchronisation and the emergence of  synchronised clusters. Indeed, the eigenvalues with their real parts close to zero (all being negative) correspond to those pseudo-normal modes which survive longer in the dynamics, and the analysis of the relevant eigenvectors allows for an understanding of the structure of such pseudo-normal modes, i.e. to single out the involved nodes. This separation of time scales is indeed a known mechanism for synchronization particularly useful in extended systems~\cite{manzano2013}.

Finally, our analysis of synchronisation of chiral harmonic network can pave the way to the study of new features, like the onset of synchronisation in hybrid chiral system, the emergence of multipartite entanglement in synchronised communities, the engineering of specific synchronisation patterns by a suitable design of the chiral network.

{\it  Acknowledgements.--}   SL and GMP acknowledge support by MUR under PRIN Project No. 2017 SRN-BRK QUSHIP, RZ acknowledges support from MINECO/AEI/FEDER through
projects EPheQuCS FIS2016-78010-P and the Mar\'ia de Maeztu Program for Units of Excellence in R$\&$D (MDM-2017-0711).


\section*{References}

\bibliography{synchro_bib2}

\begin{thebibliography}{10}

\bibitem{pikovsky2001}
Arkady Pikovsky, Michael Rosenblum, and J{\"u}rgen Kurths.
\newblock {\em Synchronization: {{A Universal Concept}} in {{Nonlinear
  Sciences}}}.
\newblock Cambridge {{Nonlinear Science Series}}. {Cambridge University Press},
  {Cambridge}, 2001.

\bibitem{refEckhardt}
B.~Eckhardt, E.~Ott, S.~H. Strogatz, D.~M. Abrams, and A.~McRobbie.
\newblock Modeling walker synchronization on the millennium bridge.
\newblock {\em Physical Review E}, 75:021110, 2007.

\bibitem{refStrogatz}
S.~H. Strogatz, I.~Stewart, and et. al.
\newblock Coupled oscillators and biological synchronization.
\newblock {\em Scientific American}, 269:102, 1993.

\bibitem{refAngelini}
L.~Angelini, G.~Lattanzi, R.~Maestri, D.~Marinazzo, G.~Nardulli, L.~Nitti,
  M.~Pellicoro, G.~D. Pinna, and S.~Stramaglia.
\newblock Phase shifts of synchronized oscillators and the systolic-diastolic
  blood pressure relation.
\newblock {\em Physical Review E}, 69:061923, 2004.

\bibitem{refLodi}
M.~Lodi, F.~Della~Rossa, F.~Sorrentino, and M.~Storace.
\newblock Analyzing synchronized clusters in neuron networks.
\newblock {\em Scientific Reports}, 10:16336, 2020.

\bibitem{refPardikes}
N.~A. Pardikes, J.~G. Harrison, A.~M. Shapiro, and M.~L. Forister.
\newblock Synchronous population dynamics in california butterflies explained
  by climatic forcing.
\newblock {\em Royal Society Open Science}, 4:170190, 2017.

\bibitem{neda2000}
Z.~N{\'e}da, E.~Ravasz, Y.~Brechet, T.~Vicsek, and A.-L. Barab{\'a}si.
\newblock The sound of many hands clapping.
\newblock {\em Nature}, 403(6772):849--850, February 2000.

\bibitem{Taylor2009}
Annette~F. Taylor, Mark~R. Tinsley, Fang Wang, Zhaoyang Huang, and Kenneth
  Showalter.
\newblock Dynamical {{Quorum Sensing}} and {{Synchronization}} in {{Large
  Populations}} of {{Chemical Oscillators}}.
\newblock {\em Science}, 323(5914):614--617, January 2009.

\bibitem{refAcebron}
Juan~A. Acebr\'on, L.~L. Bonilla, Conrad~J. P\'erez~Vicente, F\'elix Ritort,
  and Renato Spigler.
\newblock The kuramoto model: A simple paradigm for synchronization phenomena.
\newblock {\em Rev. Mod. Phys.}, 77:137--185, Apr 2005.

\bibitem{refMaianti}
M.~Maianti, S.~Pagliara, G.~Galimberti, and F.~Parmigiani.
\newblock Mechanics of two pendulums coupled by a stressed spring.
\newblock {\em American Journal of Physics}, 77(9):834--838, 2009.

\bibitem{refArenas}
Alex Arenas, Albert Diaz-Guilera, Jurgen Kurths, Yamir Moreno, and Changsong
  Zhou.
\newblock Synchronization in complex networks.
\newblock {\em Physics Reports}, 469(3):93--153, 2008.

\bibitem{refPantaleone}
James Pantaleone.
\newblock Synchronization of metronomes.
\newblock {\em American Journal of Physics}, 70(10):992--1000, 2002.

\bibitem{Hanggi2006}
Igor Goychuk, Jes{\'u}s {Casado-Pascual}, Manuel Morillo, J{\"o}rg Lehmann, and
  Peter H{\"a}nggi.
\newblock Quantum {{Stochastic Synchronization}}.
\newblock {\em Physical Review Letters}, 97(21):210601, November 2006.

\bibitem{Zhirov2008}
O.~V. Zhirov and D.~L. Shepelyansky.
\newblock Synchronization and {{Bistability}} of a {{Qubit Coupled}} to a
  {{Driven Dissipative Oscillator}}.
\newblock {\em Physical Review Letters}, 100(1):014101, January 2008.

\bibitem{Galve2012}
Gian~Luca Giorgi, Fernando Galve, Gonzalo Manzano, Pere Colet, and Roberta
  Zambrini.
\newblock Quantum correlations and mutual synchronization.
\newblock {\em Physical Review A}, 85(5):052101, May 2012.

\bibitem{Lee2013}
Tony~E. Lee and H.~R. Sadeghpour.
\newblock Quantum {{Synchronization}} of {{Quantum}} van der {{Pol
  Oscillators}} with {{Trapped Ions}}.
\newblock {\em Physical Review Letters}, 111(23):234101, December 2013.

\bibitem{refMendoza}
Ignacio Hermoso~de Mendoza, Leonardo~A. Pach\'on, Jes\'us G\'omez-Garde\~nes,
  and David Zueco.
\newblock Synchronization in a semiclassical kuramoto model.
\newblock {\em Phys. Rev. E}, 90:052904, Nov 2014.

\bibitem{galve2017}
Fernando Galve, Gian Luca~Giorgi, and Roberta Zambrini.
\newblock Quantum {{Correlations}} and {{Synchronization Measures}}.
\newblock In Felipe~Fernandes Fanchini, Diogo de~Oliveira Soares~Pinto, and
  Gerardo Adesso, editors, {\em Lectures on {{General Quantum Correlations}}
  and Their {{Applications}}}, Quantum {{Science}} and {{Technology}}, pages
  393--420. {Springer International Publishing}, {Cham}, 2017.

\bibitem{manzano2013}
Gonzalo Manzano, Fernando Galve, Gian~Luca Giorgi, Emilio
  {Hern{\'a}ndez-Garc{\'i}a}, and Roberta Zambrini.
\newblock Synchronization, quantum correlations and entanglement in oscillator
  networks.
\newblock {\em Scientific Reports}, 3(1):1439, March 2013.

\bibitem{cabot2018}
Albert Cabot, Fernando Galve, V{\'i}ctor~M. Egu{\'i}luz, Konstantin Klemm,
  Sabrina Maniscalco, and Roberta Zambrini.
\newblock Unveiling noiseless clusters in complex quantum networks.
\newblock {\em npj Quantum Information}, 4(1):1--9, November 2018.

\bibitem{refBellomo}
B.~Bellomo, G.~L. Giorgi, G.~M. Palma, and R.~Zambrini.
\newblock Quantum synchronization as a local signature of super- and
  subradiance.
\newblock {\em Phys. Rev. A}, 95:043807, Apr 2017.

\bibitem{refHush}
Michael~R. Hush, Weibin Li, Sam Genway, Igor Lesanovsky, and Andrew~D. Armour.
\newblock Spin correlations as a probe of quantum synchronization in
  trapped-ion phonon lasers.
\newblock {\em Phys. Rev. A}, 91:061401, Jun 2015.

\bibitem{refMilitello}
Benedetto Militello, Hiromichi Nakazato, and Anna Napoli.
\newblock Synchronizing quantum harmonic oscillators through two-level systems.
\newblock {\em Phys. Rev. A}, 96:023862, Aug 2017.

\bibitem{lodahl2017}
Peter Lodahl, Sahand Mahmoodian, S{\o}ren Stobbe, Arno Rauschenbeutel, Philipp
  Schneeweiss, J{\"u}rgen Volz, Hannes Pichler, and Peter Zoller.
\newblock Chiral quantum optics.
\newblock {\em Nature}, 541(7638):473--480, January 2017.

\bibitem{refPetersen}
Jan Petersen, J{\"u}rgen Volz, and Arno Rauschenbeutel.
\newblock Chiral nanophotonic waveguide interface based on spin-orbit
  interaction of light.
\newblock {\em Science}, 346(6205):67--71, 2014.

\bibitem{refMitsch}
R.~Mitsch, C.~Sayrin, B.~Albrecht, P.~Schneeweiss, and A.~Rauschenbeutel.
\newblock Quantum state-controlled directional spontaneous emission of photons
  into a nanophotonic waveguide.
\newblock {\em Nature Communications}, 5:5713, 2014.

\bibitem{refKornovan}
D.~F. Kornovan, M.~I. Petrov, and I.~V. Iorsh.
\newblock Transport and collective radiance in a basic quantum chiral optical
  model.
\newblock {\em Physical Review B}, 96:115162, 2017.

\bibitem{refColes}
R.~J. Coles, D.M. Price, J.E. Dixon, B~Royall, E.~Clarke, P.~Kok, M.S.
  Skolnick, A.M. Fox, and M.N. Makhonin.
\newblock Chirality of nanophotonic waveguide with embedded quantum emitter for
  unidirectional spin transfer.
\newblock {\em Nature Communications}, 7:11183, 2016.

\bibitem{gardiner1985}
C.~W. Gardiner and M.~J. Collett.
\newblock Input and output in damped quantum systems: {{Quantum}} stochastic
  differential equations and the master equation.
\newblock {\em Physical Review A}, 31(6):3761--3774, 1985.

\bibitem{gardiner1993}
C.~W. Gardiner.
\newblock Driving a quantum system with the output field from another driven
  quantum system.
\newblock {\em Physical Review Letters}, 70(15):2269--2272, April 1993.

\bibitem{carmichael1993a}
H.~J. Carmichael.
\newblock Quantum trajectory theory for cascaded open systems.
\newblock {\em Physical Review Letters}, 70(15):2273--2276, April 1993.

\bibitem{transientbookchapter}
Gian~Luca Giorgi, Albert Cabot, and Roberta Zambrini.
\newblock Transient {{Synchronization}} in {{Open Quantum Systems}}.
\newblock In Bassano Vacchini, Heinz-Peter Breuer, and Angelo Bassi, editors,
  {\em Advances in {{Open Systems}} and {{Fundamental Tests}} of {{Quantum
  Mechanics}}}, Springer {{Proceedings}} in {{Physics}}, pages 73--89, {Cham},
  2019. {Springer International Publishing}.

\bibitem{giovannetti2012a}
V.~Giovannetti and G.~M. Palma.
\newblock Master {{Equations}} for {{Correlated Quantum Channels}}.
\newblock {\em Physical Review Letters}, 108(4):040401, January 2012.

\bibitem{wang2005}
Zheng Wang and Shanhui Fan.
\newblock Optical circulators in two-dimensional magneto-optical photonic
  crystals.
\newblock {\em Optics Letters}, 30(15):1989--1991, August 2005.

\bibitem{koch2010}
Jens Koch, Andrew~A. Houck, Karyn~Le Hur, and S.~M. Girvin.
\newblock Time-reversal-symmetry breaking in circuit-{{QED}}-based photon
  lattices.
\newblock {\em Physical Review A}, 82(4):043811, October 2010.

\bibitem{kamal2011}
Archana Kamal, John Clarke, and M.~H. Devoret.
\newblock Noiseless non-reciprocity in a parametric active device.
\newblock {\em Nature Physics}, 7(4):311--315, April 2011.

\bibitem{feng2011}
Liang Feng, Maurice Ayache, Jingqing Huang, Ye-Long Xu, Ming-Hui Lu, Yan-Feng
  Chen, Yeshaiahu Fainman, and Axel Scherer.
\newblock Nonreciprocal {{Light Propagation}} in a {{Silicon Photonic
  Circuit}}.
\newblock {\em Science}, 333(6043):729--733, August 2011.

\bibitem{stannigel2012}
K.~Stannigel, P.~Rabl, and P.~Zoller.
\newblock Driven-dissipative preparation of entangled states in cascaded
  quantum-optical networks.
\newblock {\em New Journal of Physics}, 14(6):063014, June 2012.

\bibitem{Glauber_Amp}
{\em Amplifiers, Attenuators and Schrödingers Cat}, chapter~14, pages
  537--576.
\newblock John Wiley \& Sons, Ltd, 2006.

\bibitem{harocheExploring2006}
Serge Haroche and Jean-Michel Raimond.
\newblock {\em Exploring the {{Quantum}}: {{Atoms}}, {{Cavities}}, and
  {{Photons}}}.
\newblock {Oxford University Press}, August 2006.

\bibitem{breuerTheory2007}
Heinz-Peter Breuer and Francesco Petruccione.
\newblock {\em The {{Theory}} of {{Open Quantum Systems}}}.
\newblock {Oxford University Press}, January 2007.

\bibitem{gardiner2004a}
Crispin Gardiner and Peter Zoller.
\newblock {\em Quantum {{Noise}}: {{A Handbook}} of {{Markovian}} and
  {{Non}}-{{Markovian Quantum Stochastic Methods}} with {{Applications}} to
  {{Quantum Optics}}}.
\newblock Springer {{Series}} in {{Synergetics}}. {Springer-Verlag}, {Berlin
  Heidelberg}, third edition, 2004.

\bibitem{wiseman2009}
Howard~M. Wiseman and Gerard~J. Milburn.
\newblock {\em Quantum {{Measurement}} and {{Control}}}.
\newblock {Cambridge University Press}, {Cambridge}, 2009.

\bibitem{jacobs2014}
Kurt Jacobs.
\newblock {\em Quantum {{Measurement Theory}} and Its {{Applications}}}.
\newblock {Cambridge University Press}, {Cambridge}, 2014.

\bibitem{giovannetti2012}
V.~Giovannetti and G.~M. Palma.
\newblock Master equation for cascade quantum channels: A collisional approach.
\newblock {\em Journal of Physics B: Atomic, Molecular and Optical Physics},
  45(15):154003, July 2012.

\bibitem{ramos2014}
Tom{\'a}s Ramos, Hannes Pichler, Andrew~J. Daley, and Peter Zoller.
\newblock Quantum {{Spin Dimers}} from {{Chiral Dissipation}} in
  {{Cold}}-{{Atom Chains}}.
\newblock {\em Physical Review Letters}, 113(23):237203, December 2014.

\bibitem{ramos2016}
Tom{\'a}s Ramos, Beno{\^i}t Vermersch, Philipp Hauke, Hannes Pichler, and Peter
  Zoller.
\newblock Non-{{Markovian}} dynamics in chiral quantum networks with spins and
  photons.
\newblock {\em Physical Review A}, 93(6):062104, June 2016.

\bibitem{Olivares2012b}
S.~Olivares.
\newblock Quantum optics in the phase space: {{A}} tutorial on {{Gaussian}}
  states.
\newblock {\em European Physical Journal: Special Topics}, 203(1):3--24, April
  2012.

\bibitem{Roccati_2021}
Federico Roccati, Salvatore Lorenzo, G~Massimo Palma, Gabriel~T Landi, Matteo
  Brunelli, and Francesco Ciccarello.
\newblock Quantum correlations in {PT} -symmetric systems.
\newblock {\em Quantum Science and Technology}, 6(2):025005, jan 2021.

\bibitem{boccaletti2002}
S.~Boccaletti, J.~Kurths, G.~Osipov, D.~L. Valladares, and C.~S. Zhou.
\newblock The synchronization of chaotic systems.
\newblock {\em Physics Reports}, 366:1--101, August 2002.

\bibitem{giorgi2012}
Gian~Luca Giorgi, Fernando Galve, Gonzalo Manzano, Pere Colet, and Roberta
  Zambrini.
\newblock Quantum correlations and mutual synchronization.
\newblock {\em Physical Review A}, 85(5):052101, May 2012.

\bibitem{giorgi2019}
Gian~Luca Giorgi, Albert Cabot, and Roberta Zambrini.
\newblock Transient {{Synchronization}} in {{Open Quantum Systems}}.
\newblock In Bassano Vacchini, Heinz-Peter Breuer, and Angelo Bassi, editors,
  {\em Advances in {{Open Systems}} and {{Fundamental Tests}} of {{Quantum
  Mechanics}}}, Springer {{Proceedings}} in {{Physics}}, pages 73--89, {Cham},
  2019. {Springer International Publishing}.

\bibitem{ollivier2001}
Harold Ollivier and Wojciech~H. Zurek.
\newblock Quantum {{Discord}}: {{A Measure}} of the {{Quantumness}} of
  {{Correlations}}.
\newblock {\em Physical Review Letters}, 88(1):017901, December 2001.

\bibitem{modi2012}
Kavan Modi, Aharon Brodutch, Hugo Cable, Tomasz Paterek, and Vlatko Vedral.
\newblock The classical-quantum boundary for correlations: {{Discord}} and
  related measures.
\newblock {\em Reviews of Modern Physics}, 84(4):1655--1707, November 2012.

\bibitem{bera2017}
Anindita Bera, Tamoghna Das, Debasis Sadhukhan, Sudipto~Singha Roy, Aditi
  Sen(De), and Ujjwal Sen.
\newblock Quantum discord and its allies: A review of recent progress.
\newblock {\em Reports on Progress in Physics}, 81(2):024001, December 2017.

\bibitem{giorda2010}
Paolo Giorda and Matteo G.~A. Paris.
\newblock Gaussian {{Quantum Discord}}.
\newblock {\em Physical Review Letters}, 105(2):020503, July 2010.

\bibitem{adesso2010}
Gerardo Adesso and Animesh Datta.
\newblock Quantum versus {{Classical Correlations}} in {{Gaussian States}}.
\newblock {\em Physical Review Letters}, 105(3):030501, July 2010.

\bibitem{manzano2013a}
Gonzalo Manzano, Fernando Galve, and Roberta Zambrini.
\newblock Avoiding dissipation in a system of three quantum harmonic
  oscillators.
\newblock {\em Physical Review A}, 87(3):032114, March 2013.

\bibitem{cabot2019}
Albert Cabot, Gian~Luca Giorgi, Fernando Galve, and Roberta Zambrini.
\newblock Quantum {{Synchronization}} in {{Dimer Atomic Lattices}}.
\newblock {\em Physical Review Letters}, 123(2):023604, July 2019.

\bibitem{li2016}
Tan Li, Tian-Yi Bao, Yan-Lei Zhang, Chang-Ling Zou, Xu-Bo Zou, and Guang-Can
  Guo.
\newblock Long-distance synchronization of unidirectionally cascaded
  optomechanical systems.
\newblock {\em Optics Express}, 24(11):12336--12348, May 2016.

\end{thebibliography}

\end{document}